\documentclass[sigconf]{acmart}
\usepackage{todonotes}
\usepackage{fancyhdr}
\usepackage{soul}

\setcopyright{none}
\renewcommand\footnotetextcopyrightpermission[1]{}

\AtBeginDocument{%
  \providecommand\BibTeX{{%
    \normalfont B\kern-0.5em{\scshape i\kern-0.25em b}\kern-0.8em\TeX}}}

\begin{document}

\title[Keeping Fun Alive]{Keeping Fun Alive: an Experience Report on Running Online Coding Camps}

\author{Ilenia Fronza}
\email{ilenia.fronza@unibz.it}
\affiliation{%
  \institution{Free University of Bozen/Bolzano}
  \streetaddress{}
  \city{Bolzano}
  \state{}
  \country{Italy}
  \postcode{}
}

\author{Luis Corral}
\email{lrcorralv@tec.mx}
\affiliation{%
  \institution{ITESM Campus Queretaro}
  \streetaddress{}
  \city{Queretaro}
  \country{Mexico}}

\author{Xiaofeng Wang}
\email{xiaofeng.wang@unibz.it}
\affiliation{%
  \institution{Free University of Bozen/Bolzano}
  \streetaddress{}
  \city{Bolzano}
  \state{}
  \country{Italy}
  \postcode{}
}

\author{Claus Pahl}
\email{claus.pahl@unibz.it}
\affiliation{%
  \institution{Free University of Bozen/Bolzano}
  \streetaddress{}
  \city{Bolzano}
  \state{}
  \country{Italy}
  \postcode{}
}

\renewcommand{\shortauthors}{Fronza et al.}

\begin{abstract}
The outbreak of the COVID-19 pandemic prohibited radically the collocation and face-to-face interactions of participants in coding bootcamps and similar experiences, which are key characteristics that help participants to advance technical work. Several specific issues are faced and need to be solved when running online coding camps, which can achieve the same level of positive outcomes for participants. One of such issues is how to keep the same level of fun that participants obtained through physical activities and interactions in the face-to-face settings. In this paper, we report on our experience and insights gained from designing and running a fully remote coding camp that exposes high school students to Agile-based Software Engineering practices to enhance their ability to develop high-quality software. To design the online coding camp, we adapted the face-to-face version of the coding camp to keep the same ``level of fun'', i.e., adaptations aimed at increasing communication, engaging participants, and introducing fun items to reduce fatigue due to prolonged computer use, while preserving the technical curriculum that enables students to attain the learning goals originally planned. The comparison with the results of the face-to-face coding camp shows that we succeeded in keeping the fun alive in the online edition, and the participants of online camp were able to produce the results at the same level of quality in terms of product and process as in the face-to-face edition. From our experience, we synthesize lessons learned, and we sketch some guidelines for educators.
\end{abstract}



\maketitle

\footnote{This is a postprint (accepted version) of: Fronza et al. (2022): Keeping Fun Alive: an Experience Report on Running Online Coding Camps, ICSE-SEET '22, May 21--29, 2022, Pittsburgh, PA, USA
\href{https://doi.org/10.1145/3510456.3514153}{\textcolor{blue}{https://doi.org/10.1145/3510456.3514153}}}

\section{Introduction}
Coding bootcamps, hackathons, and coding-intensive learning experiences have been available for long, and they are continuously leveraged by students and practitioners to start, consolidate or deepen their knowledge on software development and enabling technologies. These learning environments are commonly space-bound, time-limited, and strongly collaborative, which makes them socially strong. Vicinity of participants, face-to-face interactions, and fun are key characteristics that help participants to advance their technical work, share best practices, and grow individually and collectively in expertise. Even though participants reported bootcamps to be more open and inclusive \cite{thayer2017barriers}, research in computing education found several barriers bootcamp participants might face. For example, stereotypes of \textit{nerdiness} and \textit{intelligence} exist \cite{thayer2017barriers} as in other computing education contexts \cite{lewis2016don}; moreover, considerable perseverance and confidence \cite{thayer2017barriers} are needed to face intensive activities.

A forced transition to remote work due to the COVID-19 outbreak brought very important learning: working in a remote setting implies not only the implementation of collaboration tools or a most effective leverage of communication channels. It requires a complete mindset of autonomy, teamwork, collaboration, technological resources, and understanding of goals. Cultivating these traits in early-career students equips them with qualifications that may enable them to mesh in a global community.
Future global workers will rely on an excellent command of these
tools and an unprecedented development of these characteristics, so
the earlier this effort is done, the easier it will be for students to become sooner highly effective global collaborators. Early exposure to
a distributed work environment will prove valuable as an effective
training field to embrace practices of online work, develop a good
command of the use of the enabling technology, and fine-tune the
human aspects that allow for remote collaboration, tremendously
necessary in the productive world of today and the future. In the process of learning how to conduct and leverage online classes, virtual events, and remote hackathons, several conditions that are unlike to appear in face-to-face settings came up. 
Since this is a rather recent issue, the existing literature is scarce 
in reporting and discussing the adjustments or adaptations needed to effectively replicate 
the traditional highly interactive, fun, face-to-face dynamics into virtual, remote, or online coding camps. 

In this paper, we report on our insights gained from designing and applying a fully remote coding camp that exposes high school students to agile-based Software Engineering practices to enhance their ability to develop high-quality software. We use as a baseline a Software Engineering-centric instructional strategy for intensive, face-to-face, project-based events for high school students \cite{fronza2020end}, in which games and fun were found to be a cornerstone of a successful outcome. Therefore, we aim at keeping the same ``level of fun'' in the online coding camp. For this reason, adaptations for the transition to online aim at increasing communication \cite{herbsleb2001global} and a sense of belonging \cite{mooney2021investigating}, engaging participants \cite{powell2021organizing}, and reducing fatigue due to prolonged computer use \cite{yousof2021possible} by proposing unplugged activities that require participants to move around and release energy before focusing again. To evaluate the success of our approach, we extend the face-to-face assessment framework \cite{fronza2020end} to understand if the online coding camp successfully emulates the development process and achieves the same quality of the developed products with respect to the face-to-face setting, while keeping fun alive.

Based on our experience, we synthesize guidelines for educators. 
The paper is organized as follows: Section \ref{sec:background} provides background and an overview of related work. Section \ref{sec:design} describes the online coding camp design adapted from the face-to-face version; Section \ref{sec:results} compares the online coding camp with the face-to-face one. Section \ref{sec:discussion} discusses the experience and synthesizes guidelines. Finally, Section \ref{sec:conclusion} concludes the paper.

\section{Background and related work}
\label{sec:background}
The body of research in Software Engineering (SE) education focuses on developing (in software engineers, computer scientists, and information technology experts) professional skills to abstract real-world problems and deliver solutions in the form of software products. Over the years, researchers proposed classroom experiences in which software processes are mapped efficiently to course sessions; moreover, they presented insightful discussions of practices, behaviors, and interactions among students \cite{liebenberg2015relevance}.  
In attempts to broaden participation in computing and engage end-users, a wide variety of outreach activities have been proposed \cite{dewitt2017we}, including intensive project-based experiences (e.g., bootcamps, hackathons, summer schools), which are increasingly popular \cite{decker2015understanding,champagne2016are} and also attract K-12 students to increase their awareness of computing. 
Radical collocation of participants and face-to-face interactions \cite{pe2018designing} are key characteristics that help participants quickly advance technical work in coding camps and similar experiences \cite{trainer2016hackathon}. For this reason,
most of the research provided guidelines to organize face-to-face events \cite{fronza2020end,gama2019developing,nandi2016hackathons,lara2016hackathons,happonen2020hackathons} and investigated their educational advantages when teaching Software Engineering concepts \cite{gama2019developing,fronza2020evaluating}, also exploring participants’ perspectives \cite{thayer2017barriers}. A recent study explored how these events can be harnessed within Software Engineering education to teach the necessary skills and competences for the students \cite{porras2018hackathons}.   

The COVID-19 pandemic made clear that flexible and resilient education systems are needed \cite{ali2020online}. The emergency prompted the organizers of coding camps and similar experiences to think whether they could be moved online, not only while waiting for a return to ``normal life'', but as a possible solution for sustainable development \cite{unesco2017education}. Although studies found no difference in student performance in conventional remote learning versus in-person learning \cite{paul2019comparative,kemp2014face,stack2015learning}, the emergency remote teaching context goes in a different direction, with students having a very low performance \cite{engzell2021learning,bidwell2020disruption}. Moreover, research in the fields of Global Software Engineering \cite{vsmite2010empirical} and computing education found several specific issues that need to be solved when moving coding camps online, 
including communication issues \cite{herbsleb2001global}, lack of a sense of belonging \cite{mooney2021investigating}, lack of engagement \cite{powell2021organizing}, and fatigue due to prolonged computer use \cite{yousof2021possible}. In addition, facilitators need to be able to replicate the face-to-face dynamics and the typical hands-on approach that dedicates little time to explaining fundamental principles in favor of example-centric and copy-paste programming \cite{hou2009cnp,porras2019code}.

From a Software Engineering process perspective, online intensive project-based experiences need to integrate different tools and strategies (tailored to the specific audience) to support the entire software development process with special attention to the phases that require strong team collaboration, such as software design \cite{jolak2020design}. Real-time collaborative code editors and compilers exist for different languages (such as CodeCollab\footnote{\url{https://codecollab.io/\#Welcome}}), but they are currently not available for block-based programming languages (BBPL) that proved to be very popular, even among small businesses and entrepreneurs. As a workaround for this problem, the BBPL \textit{App Inventor} programming may leverage \textit{AI2 Project Merger}, in which potentially two authors may produce two distinct functional areas of a project and eventually merge them in a single project\footnote{\url{https://appinventor.mit.edu/explore/resources/ai2-project-merger}}. Additionally, even though Agile methods accommodate the K-12 environment \cite{fronza2019bringing} and are suitable for development projects facing high uncertainty \cite{cockburn2002agile}, they rely heavily on face-to-face communication \cite{cockburn2001agile}. 
When coding camps go fully online, questions arise naturally regarding whether agile methods could still offer the same level of accommodation and whether they could be implemented effectively to support the development of projects and teamwork. Finally, according to the recent learning theories, students prefer to have an active role in their learning process \cite{masethe2017scoping}, and learning depends on reflection upon experiences \cite{kolb2014experiential}. In particular, games promote learning by merging educational content and entertainment activities that increase engagement, emotion, and motivation \cite{sorathia2012learning}. Research in SE education suggests as well to involve students in active learning experiences that provide \textit{takeaway messages} \cite{fronza2020end,scharlau2013games} by mimicking the "real-world" SE \cite{kurkovsky2019active}. Although games are not the silver bullet \cite{beecham2017best}, researchers proposed games to teach several SE topics, including programming \cite{heininger2017teaching}, agile concepts \cite{fronza2020end}, global software engineering \cite{vsablis2019building}, requirements engineering \cite{kurkovsky2019active}, cross-domain stakeholder-alignment \cite{koehlke2021cross}, testing \cite{lHorincz2021experience}, and software quality assurance \cite{morales2021learning}. Moreover, games have been included also in bootcamps \cite{lynch2011agile,fronza2020end}. Nevertheless, most of the proposed games have been conceived to be carried out in face-to-face settings because of their key characteristic (i.e., use of tangible objects). 

Despite all the related challenges, online intensive project-based experiences have been recently proposed to target problems related to COVID-19 \cite{vermicelli2021can,bolton2021virtual,hossain2020accelerating,gama2021online}. 
However, literature is lacking reports on actions needed to take advantage of past face-to-face experiences to organize online intensive project-based events. In particular, there are limited insights on how to keep the same level of fun to engage participants and reduce their fatigue due to prolonged computer use. This experience report intends to offer the lessons we have learned through designing and running online coding camps.

\section{Online Coding camp design}
\label{sec:design}
We use as a baseline our Software Engineering-centric instructional strategy for intensive, face-to-face project-based events for high school students \cite{fronza2020end}, which has the same target audience (i.e., highschoolers with little or no previous knowledge of software development) and goal (i.e., to expose participants to agile-based software engineering practices through the development of mobile apps). Games, activities, and team dynamics were the cornerstone of a successful face-to-face coding camp; participants consistently highlighted the importance of engaging games and activities, and mentioned this aspect among one of the reasons why they would suggest coding camp to friends \cite{fronza2020end}. Therefore, while transforming to a fully remote format, we
introduced adaptations to replicate the face-to-face dynamics and the typical hands-on approach. In particular, we aimed at ``keeping fun alive'', i.e., the main goal of the adaptations was to increase communication \cite{herbsleb2001global} and a sense of belonging \cite{mooney2021investigating}, engage participants \cite{powell2021organizing}, and reduce fatigue due to prolonged computer use \cite{yousof2021possible}. This section describes the changes with respect to the face-to-face coding camp \cite{fronza2020end}. 

\subsection{Instructional strategy}
 As in the face-to-face coding camp \cite{fronza2020end}, the online version consists of twenty hours of activity (one four-hour session for five days) divided into five sessions as shown in Table \ref{tab:timetable}.
 
 \begin{table}[h]
  \caption{Timetable of the online coding camp.}
  \label{tab:timetable}
  \begin{tabular}{ccp{5.5cm}}
    \toprule
    Session & Hours & Activities\\
    \midrule
    1 & 4  & Foundations of logical thinking, structured sequencing, and data abstraction\\
    2-4 & 12 & Iterative development of mobile apps\\
    5 & 4 & Completion and presentation\\
  \bottomrule
\end{tabular}
\end{table}

Table \ref{tab:mapping} summarizes the instructional strategy in the online setting, which keeps one of the key characteristics of the face-to-face instructional strategy (i.e., learning-by-playing): specific activities let participants reason action courses when there is a need for planning, managing, or empowering. As shown in Table \ref{tab:mapping}, each strategy fosters eXtreme Programming (XP) practices (which fit K-12 education \cite{kastl2016starting,meerbaum2010agile,fronza2019bringing}) that participants heuristically mix during their activity. While some strategies (marked with $\ast$ in Table \ref{tab:mapping}) required limited changes, new games (marked with $\ast\ast$ in Table \ref{tab:mapping}) substituted the face-to-face ones. For each new game, Table \ref{tab:mapping} shows the replaces of face-to-face games and
provides a rationale for the replacement. Online games required participants moving around and releasing energy before focusing again. After each game, 15 minutes were reserved for reflections to share thoughts and collect students' opinions on the takeaway message of each game. Thus, 20-30 minutes of every session were dedicated to games.

The remaining part of this section focuses on the changes with respect to the face-to-face coding camp (i.e., $\ast$ and $\ast\ast$ in Table \ref{tab:mapping}).

\begin{table*}[!hbt]
  \caption{Elements of the online instructional strategy (adapted from \cite{fronza2020end}) and changes with respect to the face-to-face (F2F) coding camp ($\ast$ adapted; $\ast\ast$ new).}
  \label{tab:mapping}
  \begin{tabular}{cp{2cm}cp{0.7cm}p{4cm}p{2cm}p{5.2cm}}
    \toprule
    &Strategy & Session(s) & Length (min.)&XP Practice & Replaced F2F strategy& Motivation for replacement\\
    \midrule
    &Manipulatable examples & 1-5 & \multicolumn{1}{c}{--} & User stories& \multicolumn{1}{c}{--} & \multicolumn{1}{c}{--}\\
    &Focus on problem-solving& 1-5 & \multicolumn{1}{c}{--} & Small releases, testing&\multicolumn{1}{c}{--}&\\
    &Alert without imposing & 1-5 & & Refactoring, testing&\multicolumn{1}{c}{--}&\multicolumn{1}{c}{--}\\
    $\ast$&We are here to help & 1-5 & \multicolumn{1}{c}{--} & Small releases, teamwork, on-site customer (i.e., one of the facilitators took the role of final customers, provided feedback, and refined requirements)&\multicolumn{1}{c}{--}&\multicolumn{1}{c}{--}\\
    $\ast$&Block-Based Programming & 2-5&  \multicolumn{1}{c}{--}& Continuous integration, refactoring, testing&\multicolumn{1}{c}{--}&\multicolumn{1}{c}{--}\\
    $\ast$&Teamwork & 1-5 & \multicolumn{1}{c}{--} & Collective ownership, pair programming, metaphor and coding standard&\multicolumn{1}{c}{--}& \multicolumn{1}{c}{--}\\
    $\ast\ast$ & Game: Paper tower & 2 & \multicolumn{1}{c}{18}& Prototyping and iterating, quick collaboration, simple design, teamwork& Marshmallow challenge & 1) The needed material (i.e., 20 A4 paper sheets) is more easily available at each participant location; 2) it introduces an element of fun when towers fall down \\
    $\ast\ast$ & Game: Color wheel & 3 & \multicolumn{1}{c}{15}& Simple design, teamwork, user stories & Tell me how you make toast & 1) It does not rely on manually drawing on post-its; 2) it reduces fatigue due to prolonged computer use by requiring to move around, activating physically and releasing energy before focusing again; 3) it introduces an element of fun when observing the type and amount of objects placed in the wheels\\
    $\ast\ast$ & Game: Thirty items & 4 & \multicolumn{1}{c}{15}& Prototyping and iterating, quick collaboration, teamwork & Letters with our bodies & 1) It reduces fatigue due to prolonged computer use by requiring to move around and release energy before focusing again; 2) it introduces an element of fun when observing collected objects\\
    $\ast\ast$ & Game: Boosting attention games & 3-4 & \multicolumn{1}{c}{10}&  Teamwork, simple design & & It increases engagement and fun, fosters networking, releases energy before focusing again\\
  \bottomrule
\end{tabular}
\end{table*}

\textbf{We are here to help ($\ast$).} During all the sessions, participants can ask for the facilitator's support by first showing the current release of the app and describing the solutions that they already tested. 

\textit{Changes:} Participants ask for support via the dedicated button in Zoom. The use of breakout rooms places the challenge of obtaining face time with facilitators. For this reason, we opt for a \textit{peer-led event}, i.e., advanced students serve as tutors \cite{lara2015peer,fronza2021enabling}. Each tutor is assigned three teams and visits the corresponding breakout rooms regularly: interactions regarding questions and answers are done directly, unless the technical need of the question requires the facilitator's opinion.

\textbf{Block-based programming ($\ast$).} We keep using a block-based programming environment in sessions 2-5, as it allows problem-driven learning \cite{morelli2011can} and fosters XP practices \cite{fronza2019bringing}.

\textit{Changes:} In an online setting, we can not provide the participants with mobile phones; thus, instead of App Inventor, we use Thunkable (https://thunkable.com), which builds applications both for iOS and Android, to make sure that most of the participants can use their own devices. Otherwise, live testing is also possible using the Thunkable emulator on PC. 

\textbf{Teamwork ($\ast$).} Facilitators form teams \cite{oakley2004turning} of three students attending three different school types; mixed teams include two females to prevent them from being in a minority within the team \cite{gammie2007group} and enhance collaboration \cite{takeda2014effects}. Teams choose a logo/name, define the goal of the app, and can collaborate on the same code (by sharing the editor's screen) or develop software parts individually.

\textit{Changes:} To ease communication and improve the learning experience, we encourage camera usage and explain why we are doing so in session 1 \cite{castelli2021students}. Participants are notified about this norm the week before so that they can prepare to be comfortable in front of the camera (e.g., room, clothes, etc.); moreover, \textit{show-and-tell} games (e.g., color wheel game) encourage and motivate camera usage. When not in plenary, teams work autonomously in Zoom breakout rooms.

\textbf{Paper tower ($\ast\ast$).} This activity substitutes the \textit{marshmallow challenge} \cite{wujec2010build} of the face-to-face coding camp \cite{fronza2020end}. Indeed, specific material is needed for the marshmallow challenge and it would be complicated to ensure that all participants (in different locations) have the same type of material (e.g., marshmallows of the same size). For this reason, teams compete (during session 2) in building the tallest freestanding tower in 18 minutes using 20 A4 paper sheets, which are easily available at each location. Being in an online setting, one team member builds the tower, while the others provide suggestions. The takeaway messages of this activity are analogous to the ones of the marshmallow challenge: prototyping and iterating can help achieve success, the importance of collaborating very quickly, and the value of cross-functional teams.

\begin{figure}[!ht]
\centering
\includegraphics[width=0.55\columnwidth,angle =90]{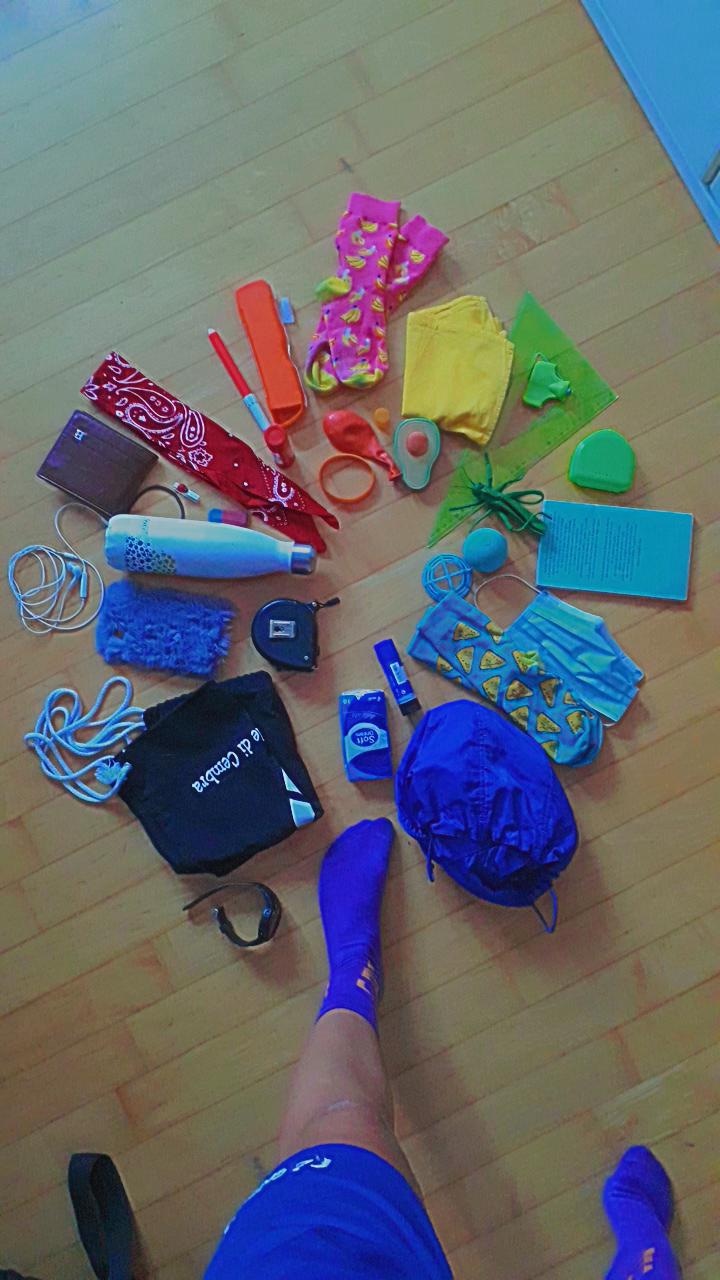}
\caption{Color wheel.}
\label{fig:wheel}
\end{figure}

\textbf{Color wheel ($\ast\ast$).} This activity substitutes the \textit{tell me how you make toast}, which heavily relies on drawing manually on post-its that are then grouped and analyzed by the team. During session 3, teams compete in creating a color wheel (Figure \ref{fig:wheel}) using the highest number of colors and objects (found in the surroundings) in 15 minutes. Being in an online setting, one team member builds the wheel while the others provide suggestions. The takeaway message of this activity is about the importance of working together toward a solution by identifying small steps.

\textbf{Thirty items ($\ast\ast$).} This activity substitutes the \textit{letters with our bodies} \cite{fronza2020end}, which is clearly not feasible in an online setting. During session 4, teams compete for 15 minutes in finding 30 items with given characteristics (e.g., shining, useless, broken) at the locations of all team members. The takeaway message of this activity is about the importance of understanding ambiguous requirements (e.g., is it possible to consider an object valid for more than one category?), and team self-organization with little or no guidance from facilitators.

\textbf{Boosting attention games ($\ast\ast$).} At the beginning of sessions 3 and 4, the first 10 minutes are dedicated to two games to start focusing: 1) \textit{who likes what?} (Session 3) consists of sharing a bingo-like screen including a series of 9 boxes with items written within. These items refer to hobbies, activities, sports, and entertainment items to choose from. The facilitator reads them out loud (for instance, ``Who likes football?'') so that participants can add themselves a mark on the box if they feel like they appreciate the mentioned item. This activity has been introduced in the online setting, as it is valuable for networking, for participants to connect with other members with similar preferences. 2) \textit{Gimme five} (Session 4). In this activity, we mention the order in which students' screens are sorted in the facilitator screen. For example, ``In the first row, we have left to right, John, Maria, and Joe; in the second row we have Bob, Stella, and Tanya''. After mentioning this arrangement, we encourage students to ``high-five'' the persons right next to them according to their screen arrangement. It is amusing and requires a big deal of coordination to have all students high-five each other. In the online setting, these games also aim at increasing engagement and fostering teamwork. 

\subsection{Participants}
Similar to the face-to-face version, the online coding camp targets high school students (aged 15-19) attending different schools (from non-vocational to computer science), i.e., having diversified disciplinary background. Moreover, the participants have little or no previous software development knowledge.

\subsection{Assessment framework}
The goal of the assessment framework is understanding if the online coding camp successfully emulates the development process and achieves the same quality of the developed products with respect to the face-to-face setting, while keeping fun alive. Therefore, we extended the framework proposed in \cite{fronza2020end}, which included product and process assessment, by including fun assessment. Under consideration of the underlying principles of Project-Based Learning \cite{romeike2012agile}, during our coding camps we limit the handing out of test/questionnaires and we prefer critique and revision, supported by observation and code inspections \cite{fronza2020end}.

\textbf{Fun assessment.}
Research has demonstrated that fun increases engagement with learning activities \cite{vieira2017assessment} and has positive effects on learning outcomes \cite{chan2019interactivity}. Despite this, the concept behind the term is not always clearly defined; moreover, there is a lack of reliable measurement tools, especially for adolescents \cite{tisza2021funq}. To understand better the reach and impact that fun activities (i.e., games) had on the participants, the facilitators organized two types of reflection sessions to collect information from participants about perceived fun:
\begin{enumerate}
    \item upon completion of each game, there was always time dedicated to reflecting about what could be important lessons that such activity can bring to the professional software development process;
    \item at the end of the coding camp, focus group interviews \cite{creswell2017research} of around 30 minutes, involving participants and student tutors separately, served to elicit views and opinions to complement the observations collected during the activities and reflection moments.
\end{enumerate}

After asking questions, the facilitators only took notes while encouraging the students to express their opinions. Reflection sessions served to explore how participants connected games to their learning process and how they considered games helpful to reduce fatigue, increase engagement, and \textit{keep fun alive} in the online setting. Student tutors did participate in the previous year's face-to-face coding camp. Therefore, besides reporting how much fun they perceived while observing the participants, they could compare the two editions in terms of fun.

Thematic analysis \cite{creswell2017research} was conducted on the collected notes based on the factors proposed by Tisza and Markopoulosto to measure adolescents' fun \cite{tisza2021funq}, i.e., answers were coded as \textit{fun} when they mentioned specific concepts (such as, curiosity, flying time, new friends, doing something new) and as \textit{not fun} with opposite concepts (such as, feeling bad, angry, sad, or forced to participate). 

{\bf Product assessment.} As we use Thunkable instead of App Inventor, we adapted the framework in \cite{fronza2020end} to extract five groups of metrics to analyze Thunkable projects from a Software Engineering perspective, namely: component metrics, computational concepts blocks, code smells, complexity metrics, and size (Table \ref{tab:ProductAssessment}). 

\begin{table}[h]
  \caption{Metrics for product assessment (adapted from \cite{fronza2020end}).}
  \label{tab:ProductAssessment}
  \begin{tabular}{p{1.7cm}p{6cm}}
    \toprule
    Group & Metric\\
    \midrule
    Component metrics & Number of components by functionality based on the categories in the Thunkable palette: authentication, data, image, layout, screens, sensors, user interface\\
        & Total Number of Components (TNC): the sum of all the components by functionality\\
        & Total Number Of Unique Blocks (NOUB)\cite{xie2015measuring}: length of the distinct list of blocks\\ 
        Computational concepts & Count of six types of blocks: conditional, function, list, logic, loop, and variable blocks\\
        Complexity & Cyclomatic Complexity (CC): number of decision points in the code plus one\\
        Size& Number of Logical Lines Of Code (LLOC) \\
        Code smells & Component names \cite{waite2017smelly}; Superfluous stuff \cite{waite2017smelly}; duplication \cite{waite2017smelly,hermans2016code}; long method \cite{waite2017smelly,hermans2016code}; Meaningful variable names \cite{grover2017tackling}\\
  \bottomrule
\end{tabular}
\end{table}

{\bf Process assessment.} Our framework capitalizes on the one proposed in \cite{fronza2020end}, as it focuses on observations of process-relevant traits focusing on XP practices. However, in the online setting, the use of breakout rooms places the challenge for facilitators of directly observing participants' behavior. To overcome it, we developed a protocol to collect observations (Table \ref{tab:ProcessAssessment}) so that we could train student tutors and delegate them to observe the assigned teams (on the second and last day of the event) by using the protocol (in a Google Form). 

\begin{table}[!h]
  \caption{Process assessment protocol.}
  \label{tab:ProcessAssessment}
  \begin{tabular}{p{5.3cm}p{2.5cm}}
    \toprule
    Observed behavior & Corresponding XP practice\\
    \midrule
    1. The team uses paper/digital sketches to drive the development of the app [never / sometimes / often] & User stories and metaphor \\
    2. The team gets ready to present a prototype at the end of each iteration [never / sometimes / often] & Small releases and iterations \\
    3. The team tests frequently [never / sometimes / often] & Refactoring/testing\\
    4. The entire team knows and can modify the code [never / sometimes / often] & Collective ownership\\
    5. Two/three team members work together on same piece of code [never / sometimes / often]  & Pair programming\\
    6. The team takes advantage of meetings with customers to get feedback [never / sometimes / often] & On-site customer\\
  \bottomrule
\end{tabular}
\end{table}

Based on the questions in the observation protocol, the student tutors observed the teams continuously and reported the results of their observations to the facilitators. Using the collected data, the facilitators could identify any critical issues and support the student tutors to manage them. In doing so, we turned the challenge into the opportunity that allowed us to understand better participants’ behavior
through analyzing the information that could not be gathered easily in the face-to-face version.

\section{Face-to-face vs. online coding camp}
\label{sec:results}
This section first assesses whether the online camp kept alive the fun that participants typically obtained in the face-to-face edition, then compares the achieved results in terms of quality of product and process in the two editions. The comparison allows us to understand if our effort of replicating fun and engagement in the online version produced desired results, which are reflected in
the quality of product and process.

The coding camp hosted 80 participants (14 F, 66 M) from ten different high school types (computer science, scientific, vocational, and non-vocational), i.e., participants had diversified disciplinary backgrounds. The participants were aged 15-19 and had little or no previous software development knowledge. In total, 27 teams that were tutored by a group of 15 second/third-year students (7 F, 8 M) who attended the previous year’s face-to-face
coding camp. Each tutor was assigned one or two teams. 

The activities started on Monday afternoon and concluded on Friday afternoon, four hours per day. Two authors of this paper facilitated all the sessions and also the face-to-face version described in \cite{fronza2020end}, which hosted 28 participants (6 F, 22M) having similar characteristics and organized in 10 teams. 

\textbf{Fun assessment.} 
During reflection sessions, the participants never reported not-fun factors (e.g., boredom or sadness \cite{tisza2021funq}). Instead, they frequently mentioned fun and explained how \textit{fun activities} changed the context and dynamics of the course (\textit{``I wondered how to survive four hours of sitting in front of the screen. Instead, time passed more quickly than I thought''}; \textit{``Games helped us socialize with people we did not know at first, which is definitely more difficult in an online context''}), assuring moments of fun (\textit{``At the end of the thirty items game, my jaws were aching from laughing''}) and relaxation (\textit{``Thanks to these activities I had to run all over my apartment, which helped stretching my legs and resting my eyes''}) . On top of that, the students commented that even though fun activities can be perceived upfront as unrelated eventually, they were able to find a connection and eventual learning that does relate to the software development process (\textit{``At first, I wondered why I had to do parkour to find objects [...] Then, I realized that deciding who had to run around to look for what was a great team game!''; \textit{``When we were developing our app, we often thought about the paper tower metaphor''}}).

According to the observations collected by the facilitators and tutors, the participants' engagement was a clear indicator
of fun. Indeed, they all did their best to achieve good results; to mention one example, some participants went all the way to the kitchen to find carrots to increase the number of orange objects in the color wheel. Moreover, the \textit{camaraderie} effect was pronounced and led to the different teams playfully competing against each other. The student tutors confirmed that they considered fun to be an essential ingredient of the face-to-face coding camp; for this reason, they were initially concerned that the online experience would be very different for new participants. However, they confirmed that they could observe fun during the coding camp and that the games played a crucial role in maintaining the participants' fun and engagement.

\begin{figure}[!ht]
\centering
\includegraphics[width=\columnwidth]{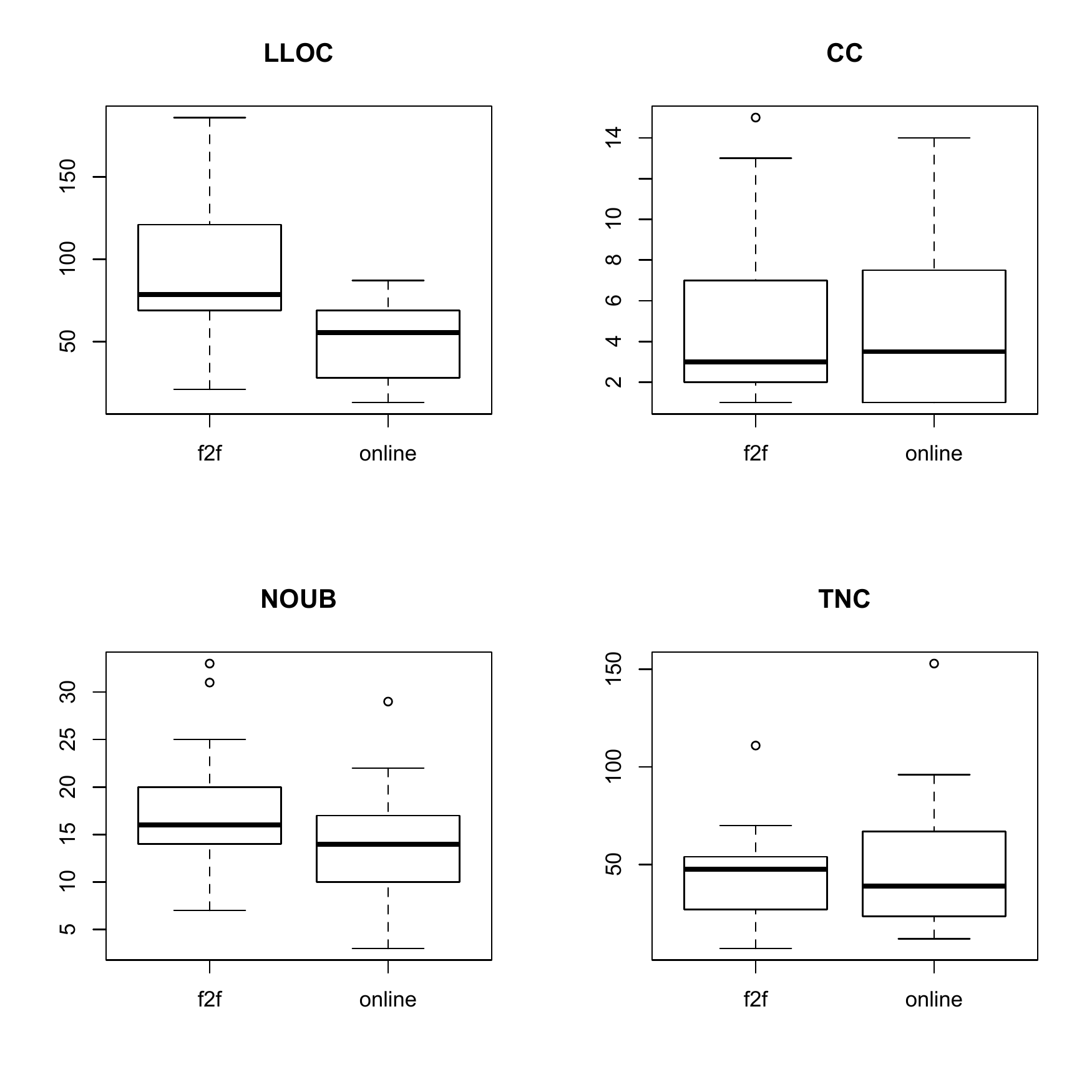}
\caption{Size and complexity: face-to-face vs. online coding camp.}
\label{fig:dimcomp}
\end{figure}

\textbf{Product assessment.} 
Figure \ref{fig:dimcomp} compares size and complexity of the projects developed at the online coding camp (27 projects in total, one per team) with those of the face-to-face edition (10 projects \cite{fronza2020end}). The only noticeable difference is represented by project size (LLOC) being slightly lower in the online edition, but still comparable to the face-to-face edition. This leads to the consideration that the capacity of the teams in terms of size and complexity of the product does not decay because of the variation in the course delivery channel. 

\begin{figure}[!ht]
\centering
\includegraphics[width=\columnwidth]{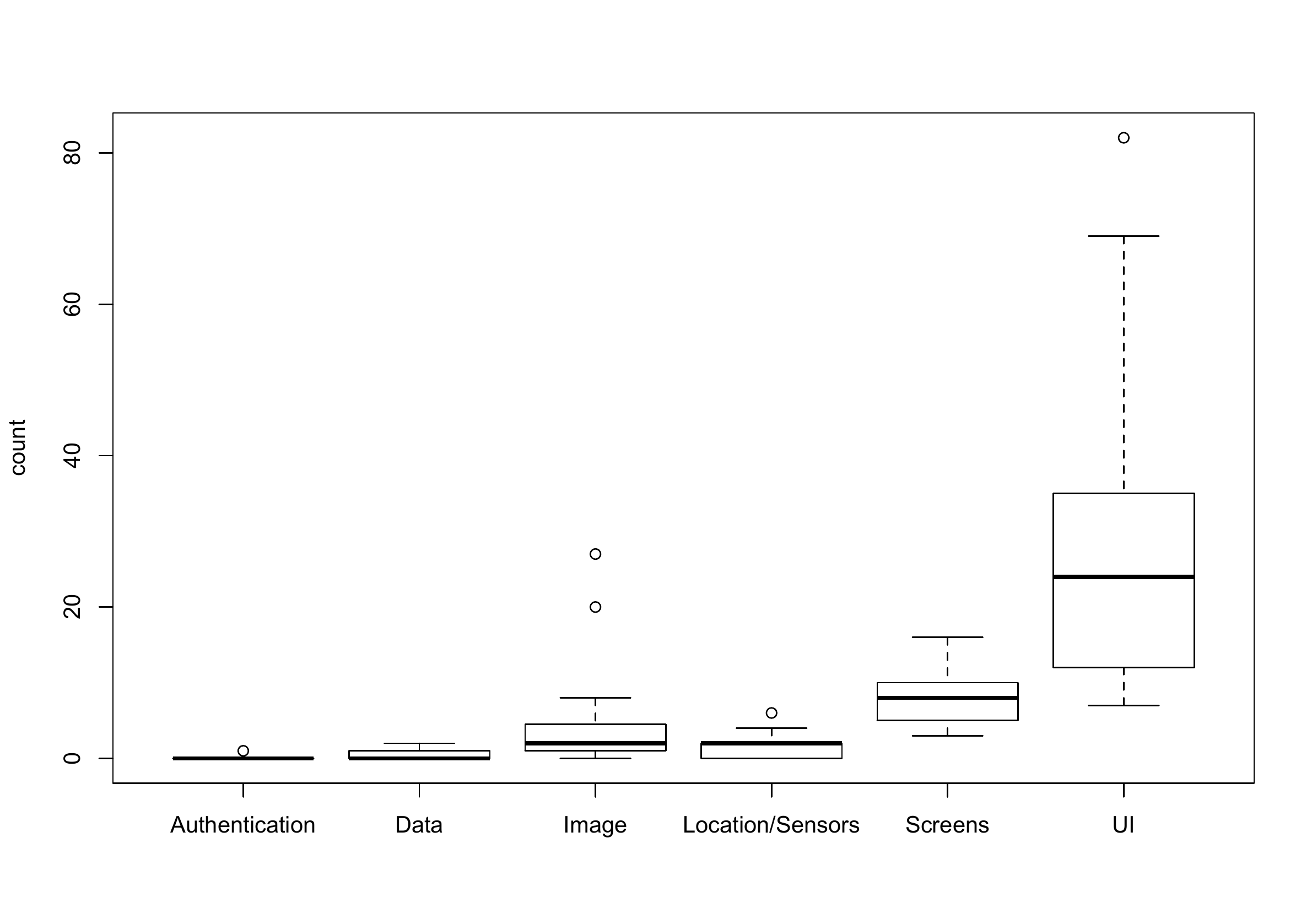}
\caption{Online coding camp: components per type.}
\label{fig:components}
\end{figure}

Figure \ref{fig:components} shows the presence of \textit{components} (based on the categories in the Thunkable palette) in the projects
of the online coding camp. User Interface (UI) components are the most present, while other components (such locations/sensors and authentication) are used less frequently. This result is similar to what happened in the products developed face-to-face \cite{fronza2020end}. 

\begin{figure}[!ht]
\centering
\includegraphics[width=\columnwidth]{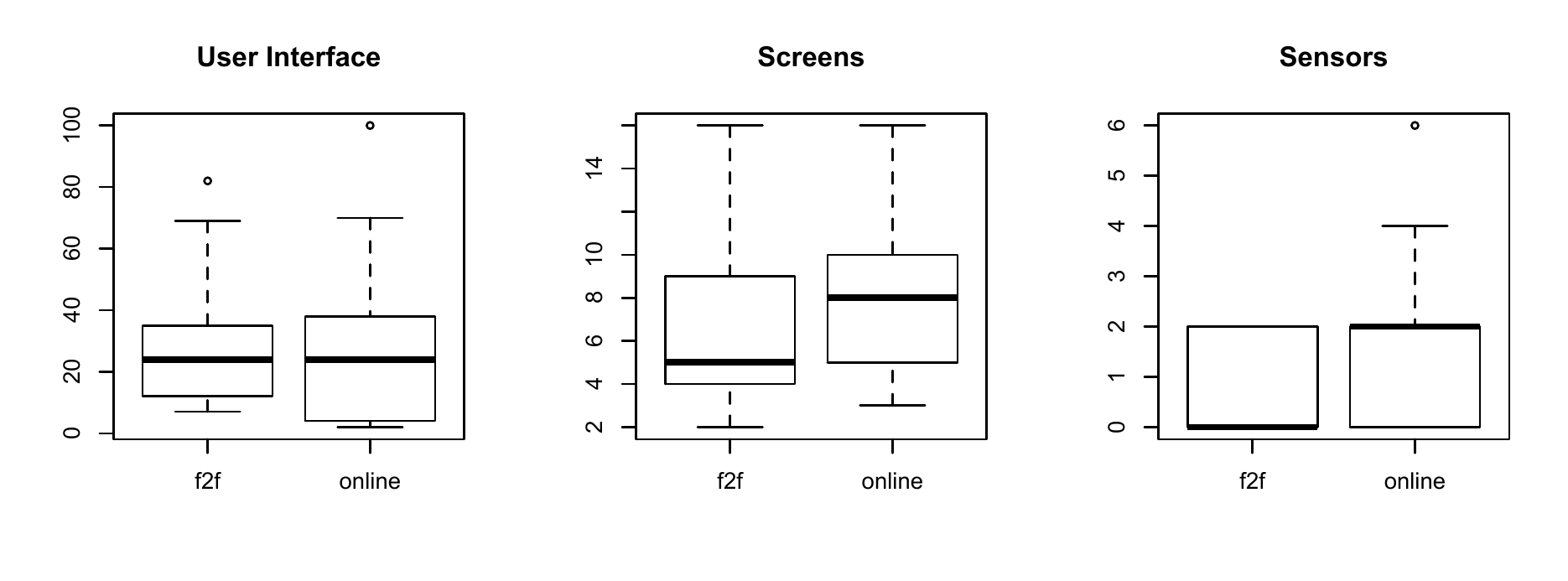}
\caption{Components: face-to-face vs. online coding camp.}
\label{fig:componentscomparison}
\end{figure}

Figure \ref{fig:componentscomparison} compares the number of components in the two editions of the coding camp (i.e., face-to-face and online). We do not compare \textit{data}, which is almost absent in both editions, and \textit{authentication},  which was only present in the Thunkable palette). Moreover, although \textit{image} components are quite common in the online edition, their distribution is not compared to the face-to-face edition because the \textit{image} category was not present in App Inventor with the same features as the category in Thunkable. Figure \ref{fig:componentscomparison} shows no major variations, which were indeed not expected as we did not change the portfolio of topics offered during the online coding camp with respect to the face-to-face syllabus. 

\begin{figure}[!ht]
\centering
\includegraphics[width=\columnwidth]{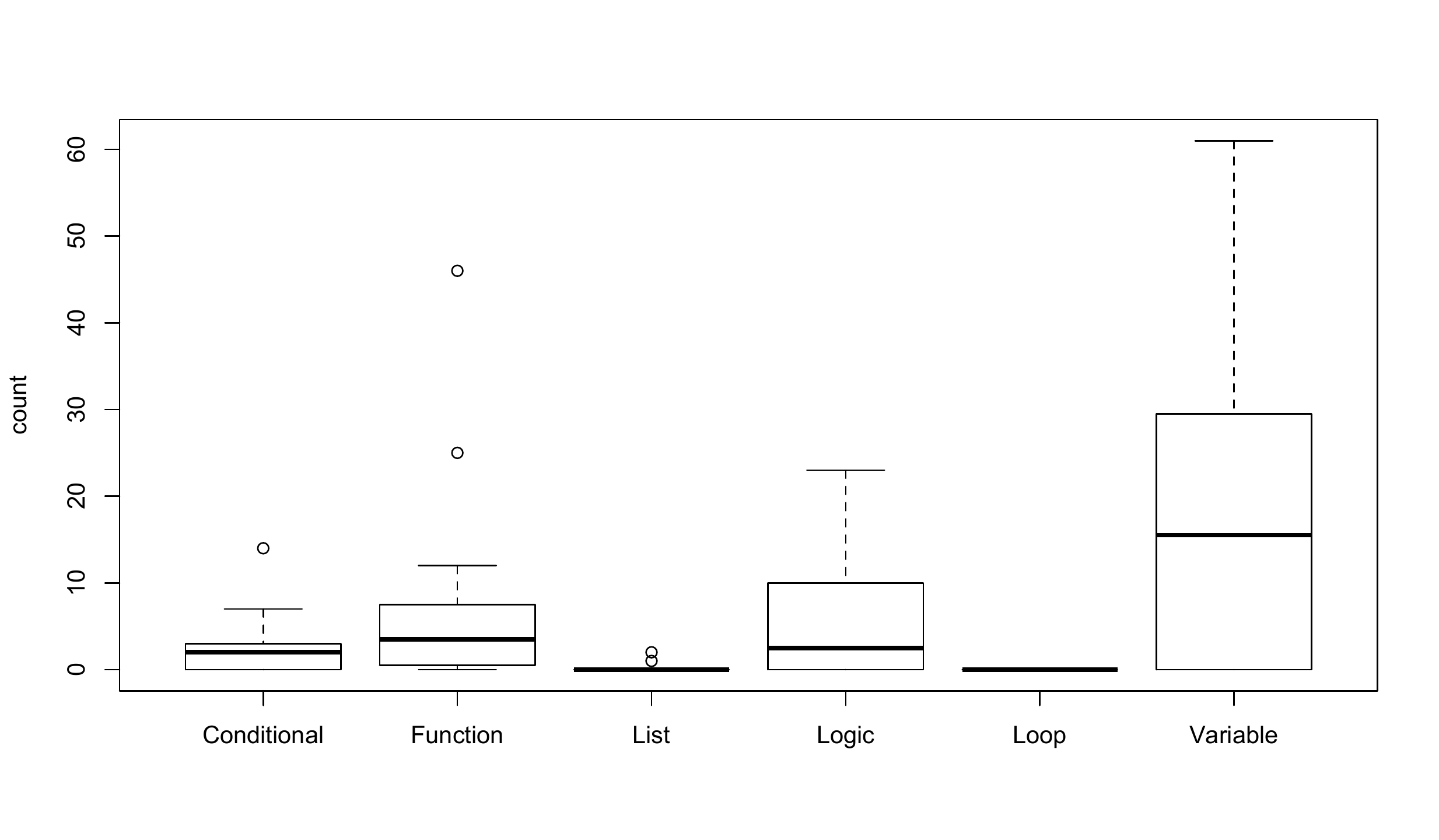}
\caption{Online coding camp: computational concept blocks.}
\label{fig:CCblocks}
\end{figure}

Figure \ref{fig:CCblocks} shows that, in the online coding camp, loops and lists are the least used \textit{computational concept blocks}, while variables, logic, and function blocks are frequently used. 

\begin{figure}[!ht]
\centering
\includegraphics[width=\columnwidth]{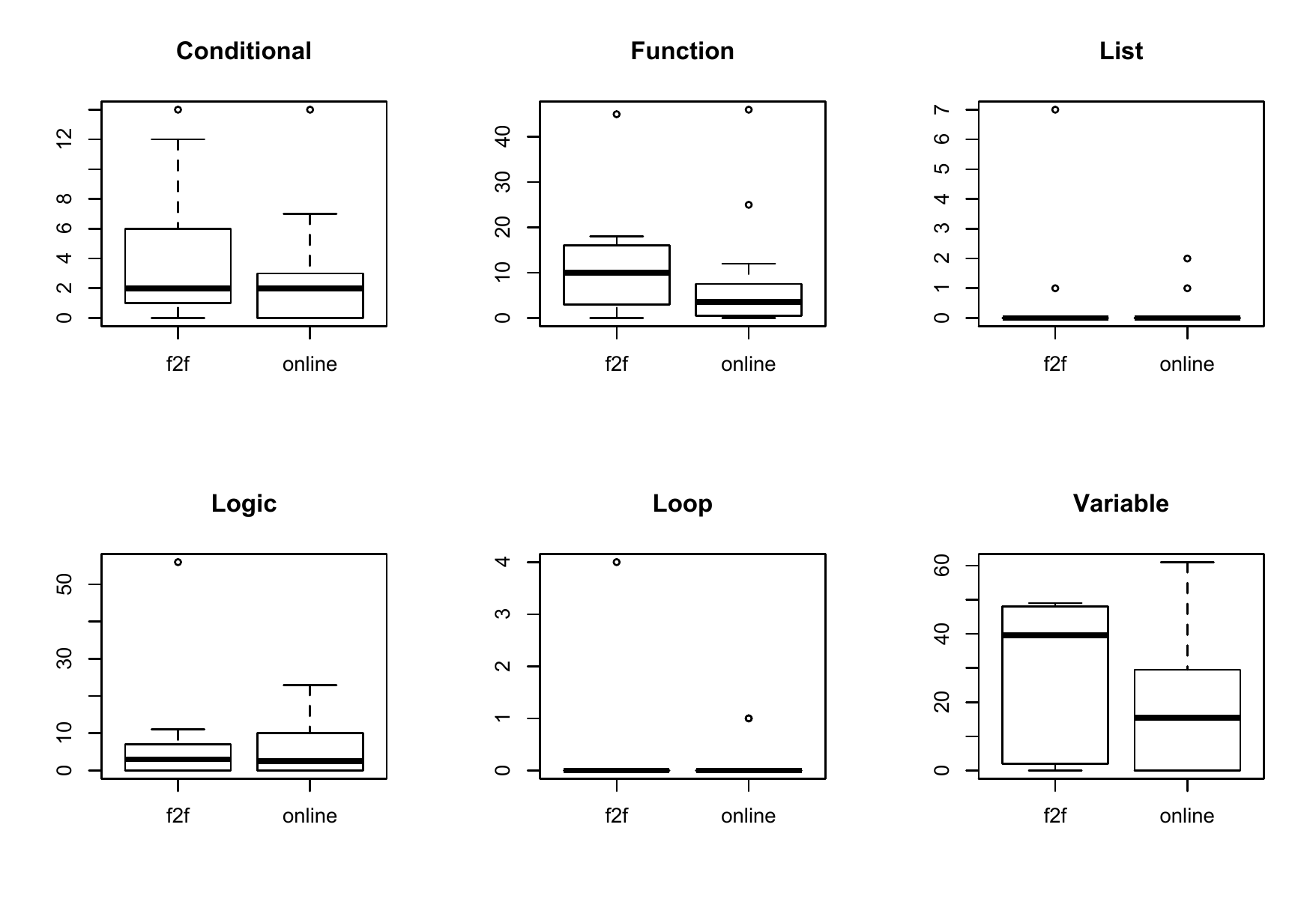}
\caption{Computational concept blocks: face-to-face vs. online coding camp.}
\label{fig:CCblockscompare}
\end{figure}

Also in this aspect, the trend offered by products developed remotely is similar to the trend observed in the face-to-face delivery method (Figure \ref{fig:CCblockscompare}). 

\begin{figure}[!ht]
\centering
\includegraphics[width=\columnwidth]{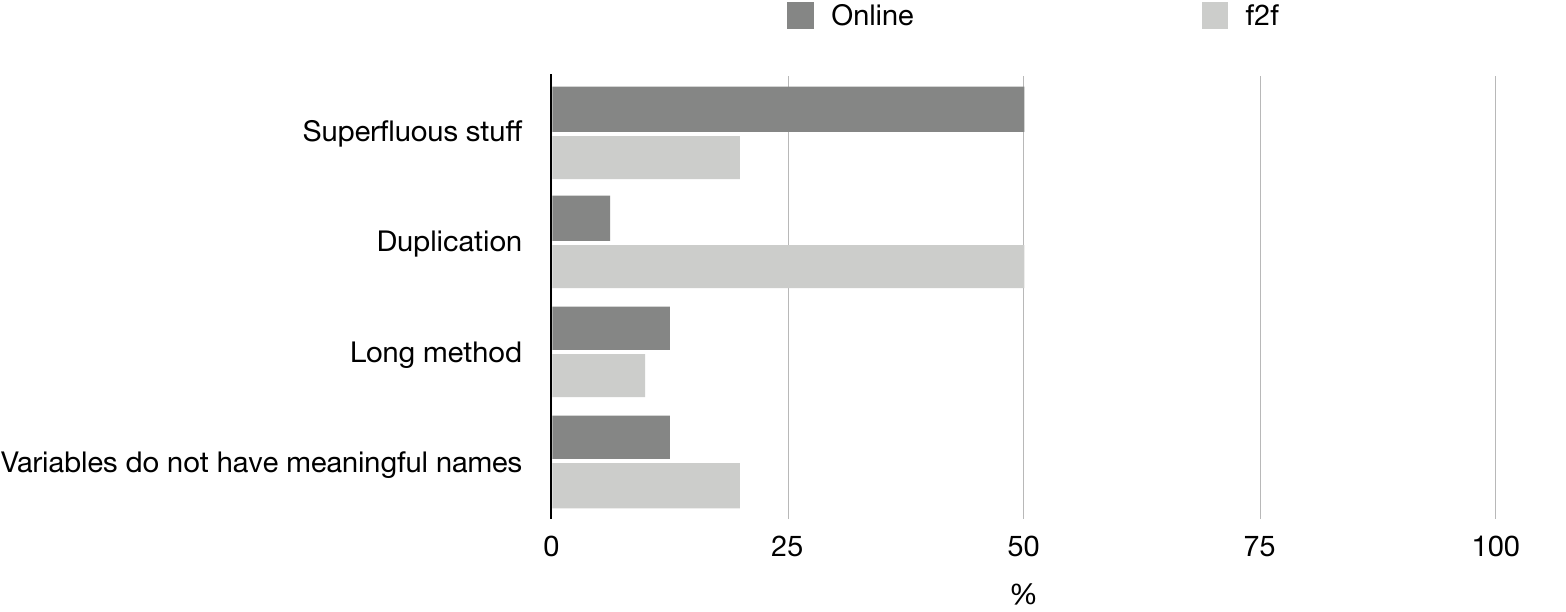}
\caption{Code smells: face-to-face vs. online coding camp.}
\label{fig:smells}
\end{figure}

The code inspection of the 27 projects developed during the online coding camp shows that half of the projects (i.e., twice the value of the face-to-face edition) suffer from the superfluous stuff code smell \textit{code smell} (Figure \ref{fig:smells}). Instead, duplication is a rather limited problem, less present with respect to the face-to-face edition. Compared to the face-to-face coding camp, an increased number of pair programming sessions have been observed, which might have contributed to reducing duplication (and to lowering project size down). Superfluous stuff in the code (i.e., discarded blocks left around) might be explained by how blocks are deleted in Thunkable: instead of dragging blocks back into the palette as in other BBPLs that participants might have already used (e.g., Scratch), it is, in fact, necessary to drag blocks to the bin in the corner; otherwise, blocks remain abandoned in the project. 

\begin{figure}[!ht]
\centering
\includegraphics[width=\columnwidth]{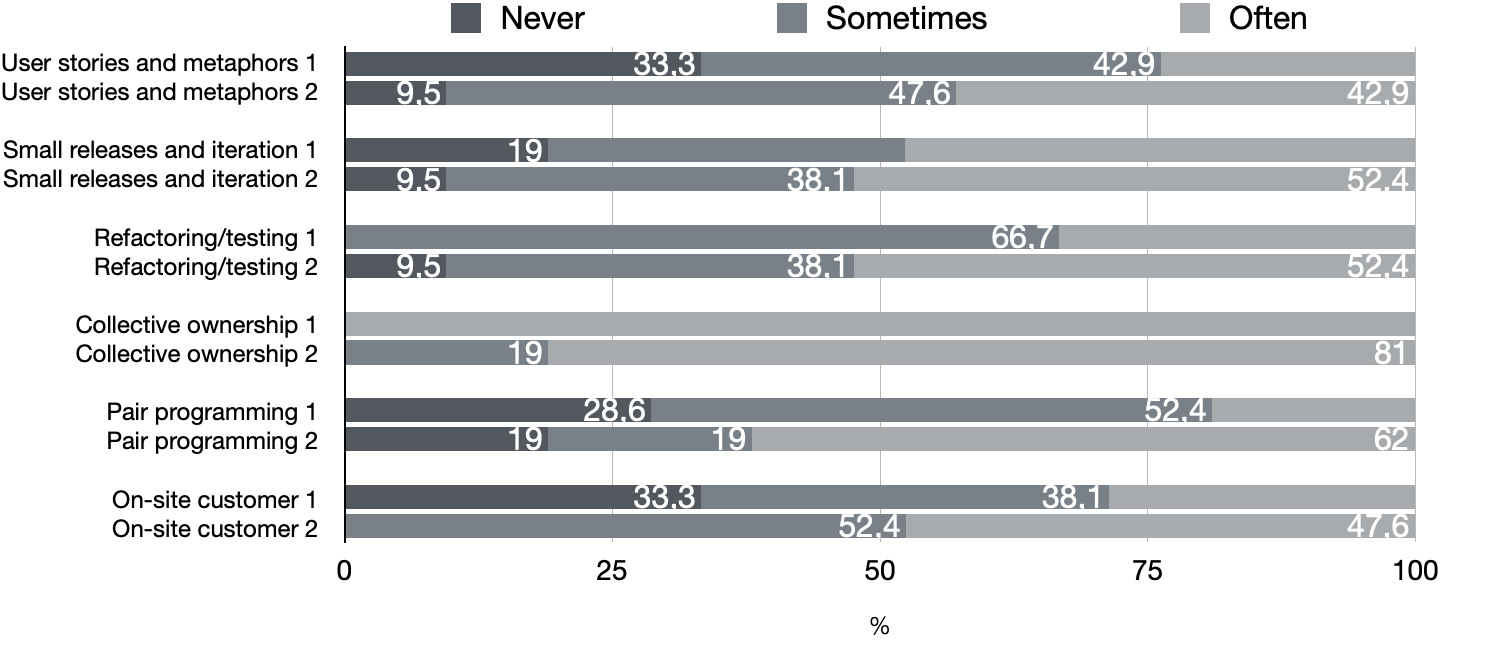}
\caption{Process assessment (beginning (1) and end (2) of the coding camp).}
\label{fig:process}
\end{figure}

\textbf{Process assessment.} Student tutors collected data on the second and last day of the coding camp using the observation protocol (Table \ref{tab:ProcessAssessment}), where each observation was an indicator of an XP practice. Figure \ref{fig:process} compares the usage of each XP practice at the two observation days.

Most of the teams used \textit{user stories and metaphors} (especially during the second part of the coding camp), i.e., they frequently used paper/digital sketches to drive app development. This could be explained by geographic distance affecting distributed design sessions \cite{jolak2020design}. \textit{Small releases and iteration} and \textit{refactoring/testing} were widely used, while the possibility to get feedback from the \textit{on-site customer} was leveraged more towards the end of the coding camp. Despite the distributed setting, the teams managed to maintain \textit{collective ownership} of the code and to work in \textit{pair programming} sessions.

Similar to the face-to-face delivery method (in which a similar observation protocol was not used), agile practices are highly promoted by facilitators during the coding camp, so observing their implementation is not surprising. However, it is important to note that students had to be more resourceful and creative in creating a proper setting to carry out some agile practices remotely (for example, pair programming sessions).

\section{Discussion}
\label{sec:discussion}
The results evidence that it is possible to emulate the development process and achieve the same quality of the developed product in both settings, while keeping fun alive in online coding camps, when the instructional goal of the camp is accompanied by support activities. Fun and relaxing items in the camp's agenda ensure not only the existence of mere ``knowledge transfer'' tasks, but also elements that activate participants promoting an overall engagement towards the technical content of the camp. The results detailed in Section \ref{sec:results} show that the effort adapting face-to-face activities enabling them to be executed fully remote supported the attainment of the desired technical results, which are observed through the quality of products developed, and the stages of the process executed, extensively discussed in Section \ref{sec:results} as well.

\subsection{Lessons Learned on the Methodological Approach}
\begin{itemize}
\item One great opportunity enabled by the online coding camp, which we were able to grasp and utilise, is that we provided student tutors with an observation protocol so that they could help us monitoring the teams' activities in their Zoom breakout rooms. It helped us understanding better participants’ behavior through analyzing the information which we could not be gathered easily in the face-to-face version, but now with online edition we could collect more easily.Before starting the coding camp, it is important to unify among student tutors the observation method and the expectations as to what results such observations should convey.
\item In the online version of the coding camp, we aimed at keeping the same type of fun generated by the activities and games in the face-to-face camp, therefore the adaptations of the games emulated the face-to-face versions and intended to keep as much as possible the physical nature of the replaced games. We demonstrated this can be done and can help to achieve the same intended results as in the face-to-face versions. An opportunity is to explore more on what ``fun'' means in online settings, and what are the online ``fun factors'',  and blend online activities with physical activities to enhance the fun that the participants could obtain, and in turn achieve better learning outcomes. This blended approach could also better attend the differences in the participants in terms of personality, background and other personal traits. Different participants can use the blended activities to achieve the same level of fun.
\item Due to the unexpected COVID situations, we had to manage to run online coding camps as quickly as possible. Therefore, it is a natural choice of adapting our existing face-to-face version and making the minimal changes possible. A missed opportunity here is to design online coding camps with a clean slate, making the best use the full potential of online tools and platforms, and design new instruction, interactivity as well as time management strategies. This is a direction worth exploring further, with fresh perspectives. 
\item In this experience report, we presented the assessment results of the online coding camp without considering the participants characteristics (e.g., diversity of previous experiences, previous courses taken, previous learn/work experience online, extroverts or introverts) which might have influenced the results, since we only knew limited background information of the participants. This reminds us that the need of finding a valid way of collecting the background information of participants so that we could inspect if participants encounter barriers \cite{thayer2017barriers} and if these barriers depend on background characteristics. 
\end{itemize}

\subsection{Lessons Learned on the Incorporation of Games and Fun activities}

Some of the insights from the introduction of fun and activating activities in an online coding camp include:

\begin{itemize}
    \item Games are essential as a strategic activity to reinforce learning: fun, engagement, move around, change context, and release energy, especially in sessions of several hours.
    \item Games must be designed in such a way as to have a take-away message that relates to the subject matter at hand (in this case, to Software Engineering).
    \item To obtain such take-away, instructors should reserve a reflection moment to share thoughts and reflections after the game. Otherwise, participants can be distracted about the real point of playing that game, or may be disappointed because they cannot show what they have done.
    \item Participating in the game and succeeding in its goal should be enabled by simple material available wherever the participant is, or found easily if notified with reasonable notice (for example the day before).
    \item When conducted virtually, instructors and student tutors should dedicate time to manage the games in breakout rooms to allow for exposure and interaction. Even if there are few rooms, participants are ``alone'' in the rooms with little or no guidance, whilst in the face-to-face version, games are played in a co-located classroom where facilitators and participants work together. Moreover, a reserved staff resource (technical facilitator) must also take care of the management of the breakout rooms so instructors can focus on mentoring.
    \item Solution sharing and reflection after the games should be run in the main Zoom room because they help keeping a connection between the teams. During the experience presented in this paper, when they later worked in breakout rooms, the teams occasionally visited other breakout rooms to help solving issues or to cross-check apps and provide comments, as what happened during the face-to-face coding camp \cite{fronza2020end}.
    \item As any context-changing activity, there is always a risk that playing a game disrupts the didactic pace of a class. By completing a totally different activity, the group should be ready to switch gears, resume the technical content and proceed. This makes the take-away speeches and the reflection discussions mentioned above especially relevant.
    
\end{itemize}    

\subsection{Lessons Learned on Remote Teamwork and Collaboration}
With regard to teamwork, collaboration and communication activities, our major insights are:

\begin{itemize}
    \item Based on the tutors' observations, the facilitators may as well intervene to encourage the group to attain a solution, or to mentor the group to explore alternative ways to find a solution.
    \item The teaching staff should give as much support and facilitation as possible to enable collaboration even in the offline aspects of the course (for example, if an exercise requires the class to draft a user interface with pencil and paper). Facilitators must inspire in this sense, using tools that the participants can replicate (such as whiteboards or paper sheets) instead of sketching or prototyping licensed tools that others may not access.
    \item  In a face-to-face setting, it is always easy to turn the head and see what students in the next table are doing. In the online edition, partial and final presentations acquire particular relevance, because the teams may lose track as to what other teams are doing, or what could be their final outcome.
\end{itemize}

\subsection{Lessons Learned on Enabling Technology}
The success of a technical camp relies much on the technology available to carry out the activities required by the course. In this regard, we observed the following learning:

\begin{itemize}
    \item The enabling technology should be selected to be executed in a common platform that can be easily supplied in any setting. For instance, working with Thunkable permits to execute a fully web-based development environment without the need of installing any software. The developed software can be tested via web, or if participants opt in, the two major mobile operating systems (iOS, Android) are fully supported by the tool. Still, minor problems commonly arise and the teaching staff should be available to mitigate eventual technical problems.
    \item A face-to-face course depends on the infrastructure provided by the course manager. A distributed, remote participation depends a lot on the technology, connectivity and infrastructure provided directly by participants. To this end, instructors should plan ahead what to do, how to deal with, and advise others what to do if technology (internet connection, software tools) were to fail, both on the instructor's and the student's end. This plan is to be discussed by the beginning of the course, to mitigate anxiety, promote resourcefulness, and creating a technology-resilient environment for all participants.
\end{itemize}



\section{Conclusions}
\label{sec:conclusion}
In this experience report, we described the implementation of a fully-remote coding camp directed to high school students, and the evaluation of the results of the experience from the point of view of assessing the ability to develop high-quality software following an agile-based Software Engineering process while keeping the fun alive. This work yields relevant results to configure future strategies that enable students to embrace a work environment to collaborate remotely with peers elsewhere in the world, simulating and stimulating a working environment that is common in a professional setting. The results show that the participants were successful attaining the goals of the camp, by constructing a working software product utilizing block-based programming tools, yet displaying important deficiencies related to the good command of structured programming. Informally, students reported overall satisfaction on the implementation of Software Engineering practices. 

A solid technical course should consider a number of didactic goals that are accomplished via traditional knowledge transfer activities (expositions, supervised exercises, repetition, and independent work). However, we believe that context-changing, gamified activities deliver important value in the process of obtaining new knowledge, as these activities fulfill a two-fold strategy: on the one hand, they reinforce messages related to the technical track of the course, and such messages are effectively discussed in the take-away sessions. On the other hand, games allow for necessary relief when spending long sessions online, and permit physical activation that is necessary after long periods of sitting in front of a screen.

Even though not targeted by this experience report, replicability and reproducibility efforts are highly recommended to generalize the conclusions yielded by this work. The observations discussed in this work follow the technical and didactic strategy of a very specific course, so it is recommended to extend the scope of this strategy to other contexts. Moreover, we advise that instances of the reproduction of this work occur in a non-distributed setting, to determine quantitatively the impact that the co-location or the remote setting can have on groups, with measurable accounts on which specific points create a major difference.  

\section{Acknowledgments}
The MobileDev coding camp (\url{https://mobiledev.inf.unibz.it}), underlying piece of this work, was fully funded by the Free University of Bozen/Bolzano, Italy. Also, authors acknowledge the support of student tutors, whose effort was not only keystone for this work but also true inspiration for camp participants.


\bibliographystyle{ACM-Reference-Format}
  \bibliography{biblio}
  \end{document}